\documentclass[aps,preprint,nofootinbib,superscriptaddress]{revtex4}%
\usepackage{amssymb}
\usepackage{amsmath}
\usepackage{epsfig}
\usepackage{amsfonts}
\usepackage{graphicx}
\usepackage{hyperref}%
\providecommand{\U}[1]{\protect\rule{.1in}{.1in}}
\begin{document}
\title{Multi-indexed Extensions of Soliton Potential and Extended Integer Solitons of
KdV Equation}
\author{Choon-Lin Ho}
\email{jcclee@cc.nctu.edu.tw}
\affiliation{Department of Physics, Tamkang University, Tamsui 251, Taiwan, R.O.C.  }
\author{Jen-Chi Lee}
\email{jcclee@cc.nctu.edu.tw}
\affiliation{Department of Electrophysics, National Chiao-Tung University and Physics
Division, National Center for Theoretical Sciences, Hsinchu, Taiwan, R.O.C.}
\date{\today }

\begin{abstract}
We calculate infinite set of initial profiles of higher integer KdV solitons,
which are both exactly solvable for the Schrodinger equation and for the
Gel'fand-Levitan-Marchenko equation in the inverse scattering transform method
of KdV equation. The calculation of these higher integer soliton solutions is
based on the recently developed multi-indexed extensions of the reflectionless
soliton potential.

\end{abstract}
\preprint{ }
\maketitle
\tableofcontents
%

\setcounter{equation}{0}
\renewcommand{\theequation}{\arabic{section}.\arabic{equation}}%

\section{\bigskip Introduction}

One well-known class of soliton of interest in both physics and mathematics is
the nontopological soliton described by the Korteweg-de Vries (KdV) equation,
i.e.,
\begin{equation}
u_{t}-6uu_{x}+u_{xxx}=0
\end{equation}
in one space $x\in(-\infty,\infty)$ and one time $t>0$ dimension. There are
many physical systems which are weakly dispersive and weekly nonlinear that
can be well described by KdV equation. The phenomena of blood pressure waves
\cite{blood}, the intenal solitary waves in oceanography \cite{sea} obeserved
in the Andaman sea and the nonlinear electrical chains etc. are among some of
them. Thus it is of interest to solve and better understand KdV equation from
different angle and in different context. Of various approaches to solving
nonlinear partial differential equation such as the KdV equation, the method
of the inverse scattering transform(IST) \cite{GGKM,DJ} invented in 1960's is
one of the most important development on this subject. According to the method
of IST, the solution of the KdV equation is converted to the solution of two
simpler linear equations, namely, the quantum mechanical Schrodinger equation
and the Gel'fand-Levitan-Marchenko (GLM) equation \cite{GL,M}.

For soliton solutions the related Schr\"{o}dinger equation is connected with
reflectionless potentials \cite{Landau}. For such reflectionless potentials,
the reflection amplitudes of the scattering states vanish, and the
corresponding GLM equation is easy to solve. One gets $2N$ continuous
parameters, $N$ norming constants $c_{n}(0)$ and $N$ energy parameters
$\kappa_{n}$, for the general $N$-soliton solution. Furthermore, as it turns
out, only the $\kappa_{n}$ parameters survive asymptotically as $t\rightarrow
\pm\infty$. These $N$ parameters fix the amplitudes, speeds and the relative
phases of the bumps of the solitons.

In this work we would like to point out a denumerably infinite set of higher
integer soliton solutions of the KdV equations. The initial profiles of these
solutions are related to the recently discovered exactly solvable quantum
mechanical systems \cite{OS1,Que,Que2,Canada,Canada2}, based on multiple
Darboux-Crum transformations \cite{Dar,Crum,Russ}. Such transformations can
generate new solvable quantum systems from the previous known ones using
certain polynomial type seed solutions. These seed functions are called the
virtual and pseudo virtual state wavefunctions \cite{OS1,OS2,OS3}. They were
obtained from the eigenfunctions by discrete symmetry operations or by using
the same functional forms of the eigenfunctions with their degrees higher than
the highest eigenlevel (these are called the over-shooting states). The
one-indexed \cite{YKM} and more complete multi-indexed extensions \cite{hls}
of the known quantum scattering problems \cite{KS} were recently calculated
along this line of thoughts.

The Darboux-Crum transformation in terms of the pseudo virtual state will
generate a new bound state below the original ground-state. Therefore it
generates a non-isospectral deformation. In this paper we will use pseudo
virtual state wavefunction to deform soliton potential with positive integer
parameter $h$. We will obtain an infinite number of reflectionless potentials,
which can be served as the initial profiles of integer KdV solitons in the
inverse scattering method mentioned above. Although the profiles we obtained
are not new soliton solutions, the method we adopted based on recently
developed multi-indexed extensions of the reflectionless soliton potential to
systematically generate higher integer KdV solitons is interesting and, most
importantly, mathematically simpler and more effective.

\section{Solvable Higher Integer $2$-Soliton $(\kappa_{0,}\kappa_{1})=(1,4)$}

We begin with a specific example of a solvable $2$-soliton potential, namely,
the simplest $1$-step deformed soliton potential under Darboux-Crum
transformation. The scattering data, or the bound state problem and the
scattering problem, of this potential was recently calculated in \cite{hls}.
The bound state problem and the scattering problem of the original soliton potential%

\begin{equation}
U(x)=-\frac{h(h+1)}{\cosh^{2}x}=-h(h+1)\,\mathrm{sech}^{2}x,~~h>0,~~-\infty
<x<\infty\label{u}%
\end{equation}
can be found in \cite{Landau,KS}. This potential contains finitely many bound
states%
\begin{align}
\phi_{n}(x)  &  =\frac{1}{(\cosh x)^{h-n}}P_{n}^{(h-n,h-n)}(\tanh
x)\nonumber\\
&  \sim\frac{1}{(\cosh x)^{h-n}}\text{ }_{2}F_{1}\left(  -n,2h-n+1,h-n+1,\frac
{1-\tanh x}{2}\right)  ,\nonumber\\
E_{n}  &  =-\kappa_{n}^{2}=-(h-n)^{2};~~~\text{\ \ \ \ \ \ \ \ }%
n=0,1,2,....\text{ \ },[h]^{\prime}, \label{boundwave}%
\end{align}
where $P_{n}^{(h-n,h-n)}(x)$ is the Jacobi polynomial and $_{2}F_{1}(x)$ is
the hypergeometric function ($[h]^{\prime}$ denotes the greatest integer not
exceeding and not equal to $h$).

The soliton potential contains a discrete symmetry%
\begin{equation}
h\rightarrow-(h+1),
\end{equation}
which can be used to construct the seed function%
\begin{equation}
\varphi_{v}(x)=(\cosh x)^{h+1+v}P_{\nu}^{(-h-1-v,-h-1-v)}(\tanh
x),~~v=0,1,2,3,4,..
\end{equation}
with energy%
\begin{equation}
E_{v}=-(h+1+v)^{2}.
\end{equation}
It turns out that for $v=1,3,5....$, the deformed potential contains pole at
$x=0$. For example, for $v=1,$%
\begin{equation}
U_{1}(x)=U(x)-2\frac{d^{2}}{dx^{2}}\log\varphi_{1}(x)=U(x)-\frac{2(h+1)}%
{\cosh^{2}x}+\frac{2}{\sinh^{2}x}%
\end{equation}
which contains pole at $x=0$. We note that although one can define the
asymptotic forms of the scattering state for this potential, the corresponding
bound state wavefunctions contain singularities. So for our purpose here, only
$v=2,4,6...$can be used to deform the soliton potential.

For simplicity and clarity of presentation, we will first use the seed
function for $h=1(n=0)$ and $v=2$%
\begin{equation}
\varphi_{2}(x)_{h=1}=\frac{1}{2}\cosh^{4}x(1+5\tanh^{2}x)
\end{equation}
to illustrate the calculation. The deformed potential is easily calculated to
be%
\begin{equation}
U_{2}(x)_{h=1}=U(x)-2\frac{d^{2}}{dx^{2}}\log\varphi_{2}(x)_{h=1}%
=-\frac{30(4\cosh^{4}x-8\cosh^{2}x+5)}{\cosh^{2}x(36\cosh^{4}x-60\cosh
^{2}x+25)} \label{u2}%
\end{equation}
which has no pole and no zero for the whole regime of $x$ and approaches $0$
asymptotically for $x\rightarrow\pm\infty$ as $U(x)_{h=1}$ does. Note that
$U_{2}(x=0)_{h=1}-U(x=0)_{h=1}=-28<0$, which suggests the existence of a
lowest new bound state for the deformed potential $U_{2}(x)_{h=1}$. The bound
state wavefunctions of the deformed potential Eq.(\ref{u2}) can be calculated
through the Darboux-Crum transformation to be%
\begin{equation}
\psi_{0}(x)=\phi_{0}^{\prime}-\frac{\varphi_{2}^{\prime}}{\varphi_{2}}\phi
_{0}=-5\,\mathrm{sech}x\tanh x\left(  1+\frac{2\,\mathrm{sech}^{2}x}%
{(1+5\tanh^{2}x)}\right)
\end{equation}
with energy%
\begin{equation}
E_{0}=-\kappa_{0}^{2}=-(h-n)^{2}=-1. \label{E0}%
\end{equation}
It can be easily shown that there is another bound state of the deformed
potential%
\begin{equation}
\psi_{1}(x)\sim\frac{1}{\varphi_{2}}=\frac{2}{\cosh^{4}x(1+5\tanh^{2}x)}%
\end{equation}
with a lower energy%
\begin{equation}
E_{1}=-\kappa_{1}^{2}=-(h+1+v)^{2}=-4^{2} \label{E1}%
\end{equation}
as was expected previously. The normalized wavefunctions and their asymptotic
forms can be calculated to be%
\begin{equation}
\psi_{0}(x)=\sqrt{\frac{15}{2}}\,\mathrm{sech}x\tanh x\left(  1+\frac
{2\,\mathrm{sech}^{2}x}{(1+5\tanh^{2}x)}\right)  \rightarrow\sqrt{\frac{10}%
{3}}e^{-x}\text{ as }x\rightarrow\infty, \label{c0}%
\end{equation}%
\begin{equation}
\psi_{1}(x)=\sqrt{\frac{15}{8}}\frac{2}{\cosh^{4}x(1+5\tanh^{2}x)}%
\rightarrow\sqrt{\frac{40}{3}}e^{-x}\text{ as }x\rightarrow\infty. \label{c1}%
\end{equation}
The constants%
\begin{equation}
c_{0}(0)=\sqrt{\frac{10}{3}},~~~c_{1}(0)=\sqrt{\frac{40}{3}} \label{cn}%
\end{equation}
in equations Eq.(\ref{c0}) and Eq.(\ref{c1}) are called norming constants. The
reflection amplitude of the scattering of the $M$-step ($M=1$ for the present
case) deformed soliton potential Eq.(\ref{u2}) was calculated to be \cite{hls}%
\begin{equation}
r_{D}(k)=r(k)\cdot\prod_{j=1}^{M}(-)^{j}\frac{k+i(h+v_{j}+1)}{k-i(h+v_{j}+1)},
\label{less}%
\end{equation}
where%
\begin{equation}
r(k)=\frac{\Gamma(1+h-ik)\Gamma(-h-ik)\Gamma(ik)}{\Gamma(-h)\Gamma
(1+h)\Gamma(-ik)}%
\end{equation}
is the reflection amplitude for the undeformed potential in Eq.(\ref{u}). In
view of the multiplicative form of $r_{D}(k)$, it is important to note that,
for integer $h=1,2,3....$, the scattering of the deformed potential remains
reflectionless as the undeformed potential due to the factor $\Gamma(-h)$ in
the denominator of $r(k).$

We are now ready to use the scattering data $\{\kappa_{n},c_{n},r_{D}(k)\}$ to
solve the KdV equation. For the reflectionless potential, $r_{D}(k)=0$, the
GLM equation is easy to solve, and the solution $u(x,t)$ is given by
\cite{DJ}
\begin{equation}
u(x,t)=-2\frac{d^{2}}{dx^{2}}\log(\det A), \label{ut}%
\end{equation}
where $A$ is a $N\times N$ matrix ($N\equiv h+1$) with elements $A_{mn}$ given
by
\begin{equation}
A_{mn}=\delta_{mn}+c_{n}^{2}(t)\frac{\exp-(\kappa_{m}+\kappa_{n})x}{\kappa
_{m}+\kappa_{n}};~~~m,n=0,1,2....,N-1. \label{A}%
\end{equation}
In Eq.(\ref{A}) $c_{n}(t)=c_{n}(0)\exp(4\kappa_{n}^{3}t)$ and is one of the
Gardner-Greene-Kruskal-Miura (GGKM) equations \cite{GGKM}.

For the present case, $N=h+1=2$. The integer $2$-soliton solution
corresponding to $(\kappa_{0,}\kappa_{1})=(1,4)$ can be calculated to be%
\begin{equation}
u(x,t)_{(1,4)}=-\frac{120e^{8t+2x}(e^{1024t}+e^{16x}+16e^{520t+6x}%
+30e^{512t+8x}+16e^{504t+10x})}{(3e^{520t}+3e^{10x}+5e^{512t+2x}%
+5e^{8t+8x})^{2}}. \label{1,4}%
\end{equation}
By taking $t=0$ in Eq.(\ref{1,4}), one reproduces the initial profile
$u(x,0)=U_{2}(x)_{h=1}$ calculated in Eq.(\ref{u2}). The asymptotic form of
the $(\kappa_{0,}\kappa_{1})=(1,4)$ solution is
\begin{equation}
u(x,t)_{(1,4)}\sim-2\sum_{n=0}^{N-1}\kappa_{n}^{2}\sec h^{2}\{\kappa
_{n}(x-4\kappa_{n}^{2}t)\pm\chi_{n}\},t\rightarrow\pm\infty,
\end{equation}
where%
\begin{equation}
\exp(2\chi_{n})=\prod_{%
\genfrac{}{}{0pt}{}{m=0}{{m\neq n}}%
}^{N-1}\left\vert \frac{\kappa_{n}-\kappa_{m}}{\kappa_{n}+\kappa_{m}%
}\right\vert ^{sgn(\kappa_{n}-\kappa_{m})}.
\end{equation}
Interestingly, it is seen that the asymptotic form of the solitary wave is
independent of $c_{n}(0)$ and is determined solely by the eigenvalues
$\kappa_{n}$'s. Note also that the previous integer 2-soliton solution
corresponds to $(\kappa_{0,}\kappa_{1})=(1,2)$. We stress that the general
2-soliton solution contains four continuous parameters $\kappa_{0}$,
$\kappa_{1}$, $c_{0}(0)$ and $c_{1}(0)$, and is given by Eq.(\ref{ut}) with
\begin{equation}
\det A=\left\{  1+\frac{c_{0}(t)^{2}}{2\kappa_{0}}e^{-2\kappa_{0}x}\right\}
\left\{  1+\frac{c_{1}(t)^{2}}{2\kappa_{1}}e^{-2\kappa_{1}x}\right\}
-\frac{c_{0}(t)^{2}c_{1}(t)^{2}}{(\kappa_{0}+\kappa_{1})^{2}}e^{-2(\kappa
_{0}+\kappa_{1})x}. \label{detA}%
\end{equation}
The $(\kappa_{0,}\kappa_{1})=(1,4)$ solution we obtained corresponds to
discrete parameters with values given in Eq.(\ref{E0}), Eq.(\ref{E1}) and
Eq.(\ref{cn}). The $(1,4)$ integer soliton solution, similar to the previous
$(1,2)$ solution, is exactly solvable quantum mechanically. On the other hand,
the scattering data obtained from, for example, $h=\frac{1}{2}$ $(n=0)$ and
$v=2$ is exactly solvable quantum mechanically, but the corresponding GLM
equation is not solvable since the reflection amplitude is not zero. It is
interesting to see that the calculation of these higher integer soliton
solutions such as the $(1,4)$ integer soliton is based on the recently
developed multi-indexed extensions of the reflectionless soliton potential.

\section{Solvable Higher Integer N-solitons}

The result of section II can be generalized to higher solvable $N$-soliton
cases (solvable in the sense of inverse scattering method). Here we present
the result for $1$-step deformation and take $v=2$, $h=1,2,3,4...$The
normalized bound state wavefunctions of the deformed potential%
\begin{equation}
U_{2}(x)_{h}=U(x)-2\frac{d^{2}}{dx^{2}}\log\varphi_{2}(x) \label{U2}%
\end{equation}
can be calculated through the Darboux-Crum transformation to be%
\begin{align}
\psi_{n}(x)  &  =\frac{1}{\sqrt{B_{n}(E_{n}-E_{h})}}\left(  \phi_{n}^{\prime
}-\frac{\varphi_{2}^{\prime}}{\varphi_{2}}\phi_{n}\right) \nonumber\\
&  =\frac{1}{\sqrt{B_{n}(E_{n}-E_{h})}}\left\{  \frac{2h-n+1}{2(\cosh
x)^{h-n+2}}P_{n-1}^{(h-n+1,h-n+1)}(\tanh x)\right. \nonumber\\
&  -\left(  \frac{(2h-n+3)\tanh x}{(\cosh x)^{h-n}}+\frac{2(2h+3)\tanh
x}{[1+(2h+3)\tanh^{2}x](\cosh x)^{h-n+2}}\right) \nonumber\\
&  \left.  \times P_{n}^{(h-n,h-n)}(\tanh x)\right\}  ,\label{bound}\\
B_{n}  &  =\frac{2^{2(h-n)}\Gamma(h+1)^{2}}{n!(h-n)\Gamma(2h-n+1)}%
\end{align}
with energy%
\begin{equation}
E_{n}=-\kappa_{n}^{2}=-(h-n)^{2};n=0,1,2...,h-1.
\end{equation}
In addition, there is an newly added bound state, given by $1/\varphi_{2}$.
The normalized form of this state is%
\begin{equation}
\psi_{h}(x)=\sqrt{\frac{2\Gamma(h+\frac{5}{2})}{\pi^{1/2}\Gamma(h+2)}}\frac
{1}{(\cosh x)^{h+3}[1+(2h+3)\tanh^{2}x]}%
\end{equation}
with lowest energy%
\begin{equation}
E_{h}=-\kappa_{h}^{2}=-(h+1+v)^{2}=-(h+3)^{2}. \label{pole}%
\end{equation}
By Eq.(\ref{less}) the scattering of the deformed potential is reflectionless.
The scattering data needed are
\begin{align}
c_{n}(0)  &  =\frac{1}{(h-n)!}\sqrt{\frac{(h-n)(2h-n+3)(2h-n)!}{(n+3)n!}%
},n=0,1,2....,h-1,\\
c_{h}(0)  &  =\frac{2^{h+2}}{h+2}\sqrt{\frac{2\Gamma(h+\frac{5}{2})}{\pi
^{1/2}\Gamma(h+2)}};\\
\kappa_{n}  &  =(h-n),n=0,1,2....,h-1,\\
\kappa_{h}  &  =h+3.
\end{align}
The general formula for the extended soliton solutions $u(x,t)$ is then
obtained by Eq.(\ref{ut}) and Eq.(\ref{A}) with $N=h+1$. By taking $t=0$ in
Eq.(\ref{ut}), one reproduces the initial profile $U_{2}(x)_{h}$ calculated in
Eq.(\ref{U2})%
\begin{equation}
U_{2}(x)_{h}=-2\frac{d^{2}}{dx^{2}}\log(\det A)_{t=0}. \label{profile}%
\end{equation}

The profile for $h=2$, for example, is the extended solvable $3$-soliton
$(1,2,5)$%
\begin{equation}
U_{2}(x)_{h=2}=u(x,0)_{(1,2,5)}=-\frac{4(144\cosh^{4}x-280\cosh^{2}%
x+147)}{\cosh^{2}x(64\cosh^{4}x-112\cosh^{2}x+49)},
\end{equation}
and
\begin{align}
u(x,t)_{(1,2,5)}  &  =-(16e^{8t+2x}(9e^{2128t}+9e^{28x}+1575e^{16(63t+x)}%
+882e^{16(66t+x)}+3252e^{14(76t+x)}\nonumber\\
&  +175e^{8(142t+x)}+49e^{8(250t+x)}+126e^{4(516t+x)}+56e^{2072t+2x}%
+126e^{2056t+6x}\nonumber\\
&  +1008e^{1128t+10x}+882e^{1072t+12x}+1575e^{1120t+12x}+1008e^{1000t+18x}%
+49e^{128t+20x}\nonumber\\
&  +175e^{992t+20x}+126e^{72t+22x}+126e^{64t+24x}+56e^{56t+26x})\nonumber\\
&  /(2e^{1072t}+2e^{16x}+14e^{4(252t+x)}+9e^{2(532t+x)}+7e^{1000t+6x}%
+7e^{72t+10x}\nonumber\\
&  +14e^{64t+12x}+9e^{8t+14x})^{2}.
\end{align}

\section{Discussion}

In this paper we have pointed out an infinite set of higher integer initial
profiles of the KdV solitons, which are both exactly solvable for the
Schrodinger equation and for the Gel'fand-Levitan-Marchenko equation in the
inverse scattering transform method of KdV equation. The calculation of these
solutions are based on the multi-indexed extensions of the reflectionless
soliton potential based on the Darboux-Crum transformation.

For simplicity and clarity of presentation, we have discussed only the case of
1-step extension using the pseudo-virtual states obtained by discrete symmetry
with integral index $v=2$. Our discussion can be straightforwardly extended to
general values of even $v$, to the general $M$-step deformations with $M=N-h$,
and to the cases using over-shooting pseudo-virtual states \cite{OS2,OS3,hls}.
Eq.(\ref{less}) ensures that the deformed potentials remain reflectionless.
For these cases, one needs to take care of the singularity problem and avoid
the singularities in the soliton profiles \cite{Russ,OS2,hls}. Thus for
extended $3$-solitons, for example, one could have two classes of solvable
solitons. The first class is%
\begin{align}
Class\text{ \ }I\text{ \ }  &  \text{: }N=3,h=2,M=1\nonumber\\
(\kappa_{0}  &  =1,\kappa_{1}=2,\kappa_{2}=v_{1}+3)\nonumber\\
v_{1}  &  =2,4,6,8......
\end{align}
and the second class is%
\begin{align}
Class\text{ \ }II\text{ \ }\text{: }  &  N=3,h=1,M=2\nonumber\\
(\kappa_{0}  &  =1,\kappa_{1}=v_{1}+2,\kappa_{2}=v_{2}+2)\nonumber\\
v_{1}  &  =2,4,6,8......,v_{2}-v_{1}=3,5,7,9...
\end{align}
In general the initial profiles of the solvable $N=h+M$ solitons contain
integer parameters $\{h,v_{1},v_{2},...v_{M}\}$ and can be calculated as
following. The undeformed soliton potential can be written as%
\begin{equation}
-h(h+1)\,\mathrm{sech}^{2}x=-2\frac{d^{2}}{dx^{2}}\log(\det A)_{t=0}%
\end{equation}
where the functional form of $A_{t=0}$ is given by Eq.(\ref{A}) with $N=h$,
and%
\begin{equation}
\kappa_{n}=h-n,c_{n}(0)=\frac{1}{(h-n)!}\sqrt{\frac{(h-n)(2h-n)!}{n!}%
},n=0,1,2....,h-1.
\end{equation}
The M-step deformed potential can be written as \cite{OS3}%
\begin{equation}
U(x)_{deformed}=U(x)_{undeformed}-2\frac{d^{2}}{dx^{2}}\log\left\vert
W[\varphi_{v_{1}},\varphi_{v_{2}},...\varphi_{v_{M}}]\right\vert .\nonumber
\end{equation}
where $W[\varphi_{v_{1}},\varphi_{v_{2}},...\varphi_{v_{M}}]$ is the Wronskian
of the seed functions $\{\varphi_{v_{1}},\varphi_{v_{2}},...\varphi_{v_{M}}%
\}$. So the solvable deformed potentials or the initial profiles of the
solitons discussed in this paper can be written as%
\begin{align}
^{U(x)_{deformed}}  &  =u(x,0)_{\{h,v_{1},v_{2},...v_{M}\}}=-2\frac{d^{2}%
}{dx^{2}}\log(\det A)_{t=0}-2\frac{d^{2}}{dx^{2}}\log\left\vert W[\varphi
_{v_{1}},\varphi_{v_{2}},...\varphi_{v_{M}}]\right\vert \nonumber\\
&  =-2\frac{d^{2}}{dx^{2}}\log\{(\det A)_{t=0}\left\vert W\right\vert
\}=-2\frac{d^{2}}{dx^{2}}\log\{(\det\hat{A})_{t=0}\left\vert \hat
{W}\right\vert \}\nonumber\\
&  =-2\frac{d^{2}}{dx^{2}}\log\{\det(\hat{A}_{t=0}\cdot\pm\hat{W}%
)\}=-2\frac{d^{2}}{dx^{2}}\log(\det\hat{A}_{t=0}^{\prime}), \label{final}%
\end{align}
which is the generalization of Eq.(\ref{profile}). In Eq.(\ref{final}),
$\hat{A}$ and $\hat{W}$ are $N\times N$ matrices extended from lower $h\times
h$ and $M\times M$ matrices without changing the values of $det$. To obtain
$\hat{A}$ from $A$, for example, one adds $M$ unit row (and column) vectors to
$A$ matrix to get a $N\times N$ matrix $\hat{A}$. For the first row (column),
one adds $(1,0,0....,0)$; the second row (column), one adds $(0,1,0,0...,0)$
etc. Similar adding can be done for the matrix $W$ to get a $N\times N$ matrix
$\hat{W}$. Finally, the resulting $N\times N$ matrix $\hat{A}_{t=0}^{\prime}$
calculated in Eq.(\ref{final}) can be reduced without changing the value of
$det$ to the form of Eq.(\ref{A}) with some values of integer $\kappa
_{n}^{\prime}$ and real $c_{n}^{\prime}(0)$ of the corresponding extended
solvable $N$-soliton.

The existence of integer $\kappa_{n}^{\prime}$ and real $c_{n}^{\prime}(0)$ in
Eq.(\ref{final}) are guaranteed since, in our approach, we know that
$U(x)_{deformed}$ is a reflectionless solvable potential. So in this case
Eq.(\ref{ut}) and Eq.(\ref{A}) can be applied.

In the beginning of our calculation, we could have set $h=0$ and did the say
$2$-step deformation. We then end up with, for $v_{1}=2$ for example, $(3,6)$,
$(3,8)$, $(3,10)....$solitons.

\section{\bigskip Acknowledgments}

We thank Ryu Sasaki for helpful comments on this work. J.C.L. would like to
thank J.C. Shaw for discussion and S.H.Lai for assistance in numerical work.
The work of C.L.H. is supported in part by the National Science Council (NSC)
of the Republic of China under Grant NSC-102-2112-M-032-003-MY3. The work of
J.C.L. is support in part by NSC-100-2112-M-009-002-MY3, the 50 billions(NTD)
project of the Ministry of Education (Taiwan) and S.T. Yau center of NCTU, Taiwan.

After the completion of the draft, we were informed by Ryu Sasaki of reference
\cite{Kay}, which addressed the issue of reflectionless potentials from a
different context.

\end{document}